\documentclass[twocolumn,aps,floatfix,showkeys,letterpaper,superscriptaddress]{revtex4}
\usepackage[utf8]{inputenc}
\usepackage[english]{babel}
\usepackage{graphicx}
\usepackage{hyperref}

\usepackage{braket}
\usepackage{siunitx}
\usepackage{xspace}
\newcommand{\BePlus}{$^9$Be$^+$\xspace}

\usepackage{float}

\begin{document}
\title{Resolved-sideband cooling of a single \BePlus ion in a Penning trap}
	
\author{Juan~M.~Cornejo}
\email[]{cornejo-garcia@iqo.uni-hannover.de}
\affiliation{Institut für Quantenoptik, Leibniz Universität Hannover, Welfengarten 1, 30167 Hannover, Germany}
\author{Johannes~Brombacher}
\affiliation{Institut für Quantenoptik, Leibniz Universität Hannover, Welfengarten 1, 30167 Hannover, Germany}
\author{Julia~A.~Coenders}
\affiliation{Institut für Quantenoptik, Leibniz Universität Hannover, Welfengarten 1, 30167 Hannover, Germany}
\author{Moritz~von~Boehn}
\affiliation{Institut für Quantenoptik, Leibniz Universität Hannover, Welfengarten 1, 30167 Hannover, Germany}
\author{Teresa~Meiners}
\affiliation{Institut für Quantenoptik, Leibniz Universität Hannover, Welfengarten 1, 30167 Hannover, Germany}
\author{Malte~Niemann}
\affiliation{Institut für Quantenoptik, Leibniz Universität Hannover, Welfengarten 1, 30167 Hannover, Germany}
\author{Stefan Ulmer}
\affiliation{RIKEN, Ulmer Fundamental Symmetries Laboratory, 2-1 Hirosawa, Wako, Saitama 351-0198, Japan}
\affiliation{Institut für Experimentalphysik, Heinrich Heine Universität Düsseldorf, Universitätsstr. 1, 40225 Düsseldorf, Germany}
\author{Christian Ospelkaus}
\affiliation{Institut für Quantenoptik, Leibniz Universität Hannover, Welfengarten 1, 30167 Hannover, Germany}
\affiliation{Physikalisch-Technische Bundesanstalt, Bundesallee 100, 38116 Braunschweig, Germany}
	
\begin{abstract}
Manipulating individual trapped ions at the single quantum level has become standard practice in radio-frequency ion traps, enabling applications from quantum information processing to precision metrology. The key to accessing this regime is motional ground state cooling. Even though there is the potential to access a completely unexplored regime of sensitivities in fundamental physics tests, ground state cooling has not been applied in the context of Penning-trap precision measurements yet. Here we report resolved-sideband cooling of the axial mode of a single \BePlus ion in a 5 Tesla Penning trap, reducing the average phonon number to $\bar{n}_z = 0.10 \pm 0.04$. This is a key step for the implementation of quantum logic techniques in high-precision Penning-trap experiments, such as matter-antimatter comparison based tests of the fundamental CPT symmetry.
\end{abstract}
	

\maketitle

Penning ion traps use a combination of static magnetic and electric fields to confine individual particles and are used in ultra-precise atomic mass measurements, $g$-factor measurements, determinations of fundamental constants and related tests of fundamental physics. Penning-trap based tests of fundamental physics \cite{smorra350foldImprovedMeasurement2018, hannekeNewMeasurementElectron2008} rely on the measurement of motional and internal frequencies of the trapped particle. A strong magnetic field $B$ and a weak electrostatic potential $\phi$ define the motional and internal frequencies\cite{brownGeoniumTheoryPhysics1986}, and fundamental properties of the trapped particles can be studied by measuring frequency ratios\cite{blaumPerspectivesTestingFundamental2021,blaumPenningTrapsVersatile2010}, such that the magnetic field cancels out to lowest order\cite{smorraBASEBaryonAntibaryon2015}. In the pursuit of today's most precise measurements, the majority of significant systematic frequency shifts and their related uncertainties can be attributed to the finite energy of the confined particle. Along its trajectory, the particle interacts with imperfections in the trapping  fields\cite{ketterFirstorderPerturbativeCalculation2014}, all the while adhering to the principles of special relativity\cite{vogelTrappedIonOscillation2010}. This results in energy-dependent frequency shifts, and introduces uncertainties in the measured frequency ratios \cite{borchert16partspertrillionMeasurementAntiprotontoproton2022}. While notable progress has been made in enhancing the attainable uniformity of the technical trapping fields\cite{smorraBASEBaryonAntibaryon2015, heisseHighprecisionMassSpectrometer2019}, the particle's localization remains an area with considerable potential for further improvement. Present precision experiments exclusively rely on the coupling of the particle's motion to a sensitive cryogenic superconducting detection circuit \cite{nagahamaHighlySensitiveSuperconducting2016} that defines the detection interface to investigate the fundamental particle properties. In addition, the circuit cools and localizes the particle, implying in most cases a thermal state of motion, with an effective mean energy that is close to the physical circuit temperature of a few Kelvin. 

Ultra-high-presicion experiments using Penning traps will greatly benefit from the reduction of systematic errors offered by full motional control over atomic ions, with applications to atomic masses~\cite{blaumHighaccuracyMassSpectrometry2006} and $g$-factor measurements~\cite{smorraBASEBaryonAntibaryon2015} or related tests of fundamental physics~\cite{blaumPenningTrapsVersatile2010}. In addition, it will allow to implement quantum logic spectroscopy~\cite{schmidtSpectroscopyUsingQuantum2005}, a technique that has enabled a new class of precision measurements in radio-frequency ion traps, where manipulating individual trapped ions at the single quantum level has become standard practice~\cite{leibfriedQuantumDynamicsSingle2003}, enabling applications ranging from quantum information processing ~\cite{ciracQuantumComputationsCold1995,winelandExperimentalIssuesCoherent1998,kielpinskiArchitectureLargescaleIontrap2002} to precision metrology~\cite{ludlowOpticalAtomicClocks2015}. The key ingredient for full control is the ability to ground-state cool the motion of the particle in the trap through resolved-sideband laser cooling. Laser cooling does not only provide localization of the particle at the quantum limit; it also does so on much shorter time scales of milliseconds compared to minutes up to hours for resonators~\cite{schneiderDoubletrapMeasurementProton2017}.  The challenge is that most species of interest in Penning-trap precision measurements cannot be readily laser-cooled directly, and that the large magnetic field of the Penning trap complicates cooling significantly~\cite{goodwinResolvedSidebandLaserCooling2016,jordanGroundStateCoolingTwoDimensional2019,gutierrezTRAPSENSORFacilityOpenring2019}. The former can be addressed by implementing sympathetic cooling schemes~\cite{heinzenQuantumlimitedCoolingDetection1990, winelandExperimentalIssuesCoherent1998,brownCoupledQuantizedMechanical2011,bohmanSympatheticCoolingTrapped2021,willSympatheticCoolingSchemes2022,meinersSympatheticCoolingSingle2018,cornejoOptimizedGeometryMicro2016}, and the latter is the subject of the present work.

\begin{figure*}[t]
	\centering
	\includegraphics[width=2.05\columnwidth]{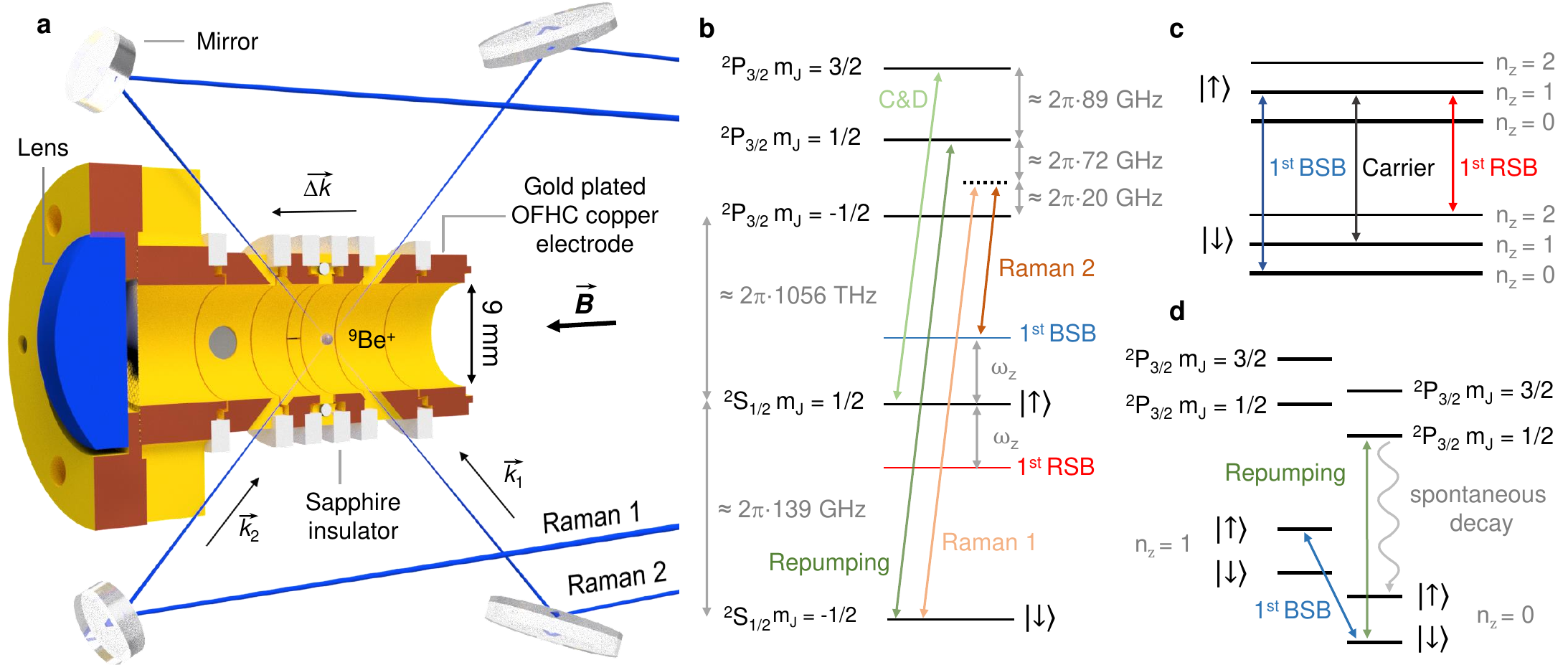}
	\caption{a, Cross-section view of the cryogenic trap. Electrodes are cylinder-shaped and made of gold-plated oxygen-free high thermal conductivity (OFHC) copper of 9~mm inner diameter, electrically insulated by sapphire rings. A lens is used to collect the fluorescence photons emitted by a single laser-cooled \BePlus ion. Four mirrors are used to guide the laser beams into the trap. The Raman laser beams are guided in a 90-degree crossed-beam configuration resulting in an effective wavevector difference along the axial direction. The cooling and repumper laser beams (not shown) use the same path as the Raman 2 laser beam. b, Relevant internal energy levels scheme of \BePlus in a 5 Tesla magnetic field. The cooling and detection (C$\&$D) as well as the repumper transition are represented by light and dark green arrows, respectively. The Raman transitions are depicted by dark and light orange for Raman lasers 1 and 2, respectively. The presence of sidebands is indicated by the red and blue levels below and above of the spin $\ket{\uparrow}$ state, respectively. c, The quantized motional energy levels of the axial mode $n_z$ = 0, $n_z$ = 1 and $n_z$ = 2 are represented for both spin states $\ket{\uparrow}$ and $\ket{\downarrow}$. The carrier and the first blue and red sidebands are depicted by black, blue and red arrows, respectively. d, Simplified internal energy level scheme represented for motional energies of $n_z$ = 1 and $n_z$ = 0. The repumper and the first blue sideband transitions are depicted by dark green and blue arrows, respectively. During the repumper pulse, the ion is re-set to the $\ket{\uparrow}$ state after spontaneous decay, which is represented by the curved gray arrow. Energy levels are not drawn to scale.}
	\label{fig:setup}
\end{figure*}

In this letter, we show ground state cooling on the axial mode of a single \BePlus ion in a 5 Tesla Penning trap by using resolved sideband cooling via a two-photon induced Raman process. Our approach directly addresses the spin-flip ``qubit'' transition within the $S_{1/2}$ ground state of  a {\em single} $^9$Be$^+$ ion, bearing the closest resemblance with wide-spread practice in radio-frequency ion traps, and is therefore complementary to the recent demonstration of ground state cooling of drumhead modes of a large \BePlus crystal~\cite{jordanGroundStateCoolingTwoDimensional2019} as well as to resolved-sideband cooling of a single $^{40}$Ca$^{+}$ ion on a forbidden quadrupole transition~\cite{goodwinResolvedSidebandLaserCooling2016}. 

Our cryogenic Penning trap system is located at the center of a superconducting magnet with a magnetic field strength $B=5\,\mathrm{T}$ to store a single \BePlus ion~\cite{niemannCryogenic9BePenning2019,cornejoOpticalStimulatedRamanSideband2023}. Among the different traps in our experiment stack, only the so-called Beryllium trap is used for laser manipulation of ions. It is depicted in Fig.~\ref{fig:setup}.a. The electric field produced by the trap electrodes in conjunction with the axial magnetic field $\vec{B}$ are used to confine the charged particle in the trap. From a classical point of view, the motion of the trapped ion is described by an axial mode with a frequency $\nu_z$ and two radial modes, the so-called magnetron and reduced cyclotron modes, with frequencies $\nu_-$ and $\nu_+$, respectively. These frequencies are related to the free cyclotron frequency $\nu_{c} = qB/(2\pi m)$ by the invariance theorem $\nu_{c}^2=\nu_+^2+\nu_z^2+\nu_-^2$~\cite{brownPrecisionSpectroscopyCharged1982}, where $q/m$ is the charge-to-mass ratio of the trapped particle.

\begin{figure}[t]
	\centering
	\includegraphics[width=1.0\columnwidth]{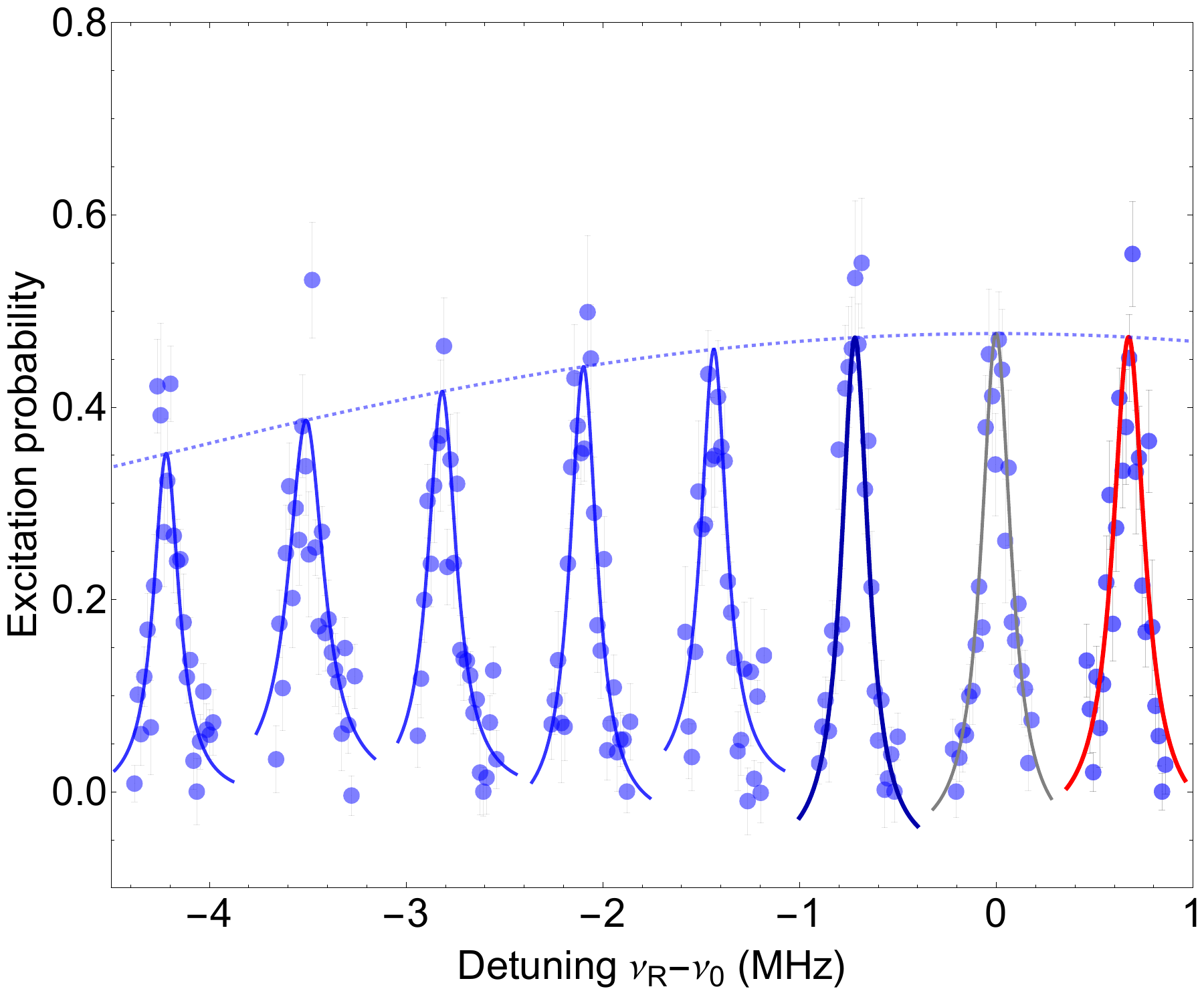}
	\caption{Resolved-sideband spectrum of a single \BePlus ion after Doppler cooling. The excitation probability is shown as a function of the Raman laser frequency difference $\nu_R$ with respect the carrier transition frequency $\nu_0$. Data points are represented by blue dots. Each data point and its respective error bar is obtained from 50 individual measurements. A Lorentzian fit to the data is represented in blue, dark blue, gray and red for the sixth to the second blue sidebands, the first blue sideband, the carrier transition and the first red sideband, respectively. The blue dashed line represents the Gaussian fit envelope to the complete sideband spectrum. }
	\label{fig:spectrum}
\end{figure}

Once a single \BePlus ion is stored in the trap~\cite{niemannCryogenic9BePenning2019, meinersFastAdiabaticTransport2023a}, Doppler cooling is performed on the ion~\cite{cornejoOpticalStimulatedRamanSideband2023}. Figure~\ref{fig:setup}.b shows the relevant energy levels for a \BePlus ion in a 5 Tesla magnetic field as well as the laser-induced transitions. A Doppler cooling and detection laser resonant with the $\ket{^2S_{1/2}, m_j=+1/2}$ $\rightarrow$ $\ket{^2P_{3/2}, m_j=+3/2}$ transition is used to cool the ion. A repumper laser resonant with the $\ket{^2S_{1/2}, m_j=-1/2}$ $\rightarrow$ $\ket{^2P_{3/2}, m_j=+1/2}$ transition prevents the accumulation of population in $\ket{^2S_{1/2}, m_j=-1/2}$. A single \BePlus ion can thus be cooled to a temperature close to the Doppler limit of $\approx0.5\,\mathrm{mK}$.
 
Two Raman laser beams with a detuning of $\approx20\,\mathrm{GHz}$ from the $\ket{\downarrow} \equiv \ket{^2S_{1/2}, m_j=-1/2}$ $\rightarrow$ $\ket{^2P_{3/2}, m_j=-1/2}$ and $\ket{\uparrow} \equiv \ket{^2S_{1/2}, m_j=+1/2}$ $\rightarrow$ $\ket{^2P_{3/2}, m_j=-1/2}$ transitions, called Raman 1 and 2, respectively, are used to drive a two-photon stimulated Raman spin-flip transition between the two ``qubit'' states~\cite{winelandExperimentalIssuesCoherent1998} $\ket{\downarrow}$ and $\ket{\uparrow}$ with an energy difference $h \cdot \nu_0\approx h \cdot 139\,\mathrm{GHz}$. In order to simultaneously address a motional mode of the trapped ion, the Raman laser beams must fulfill the resonance condition given by the effective Hamiltonian of the transition~\cite{winelandExperimentalIssuesCoherent1998,mielke139GHzUV2021,cornejoOpticalStimulatedRamanSideband2023}. For this, the Raman laser beams also need to exhibit a wavevector difference $\vec{\Delta k} = \vec{k}_1-\vec{k}_2$ with a finite projection on the axial direction in order to be able to address the axial motion as shown in Fig.~\ref{fig:setup}.a. In addition, the Raman laser beams' frequency difference $\nu_{R} = \nu_1 - \nu_2$ must satisfy the specific resonance condition  $\nu_{R}= \nu_0 + m \cdot\nu_z$, where $m$ is an integer. For $m = 0$, the Raman lasers address the carrier transition at $\nu_0$. For $m \neq 0$, a motional sideband spectrum around the carrier transition is expected. Transitions with $m>0$ ($m<0$) are known as blue (red) sideband transitions. After Doppler cooling, the ion is expected to be in a thermal state of motion. If a mean phonon number much larger than one is obtained, the motional sideband spectrum will follow a Doppler broadened envelope determined by the ion temperature~\cite{mavadiaOpticalSidebandSpectroscopy2014,mielke139GHzUV2021}. 

Before implementing sideband cooling, we need to identify the proper sideband transitions and frequencies. To that end, after Doppler cooling, we turn on the two Raman beams for a given interaction time. Afterwards, we turn on the cooling laser; if the ion is bright (scatters photons from the laser beam), it is determined to be in the $\ket{\uparrow}$ state, and in $\ket{\downarrow}$ otherwise. Following the detection, the repumper laser beam is applied to re-set the spin state to $\ket{\uparrow}$. By stepping $\nu_{R}$ and repeating the experiment, we obtain a sideband spectrum as in Figure~\ref{fig:spectrum}, where the first six blue sidebands, the carrier transition and the first red sideband transition ($m=6,\ldots,1,0,-1$) are shown for a single Doppler-cooled \BePlus ion stored in the trap at an axial frequency $\nu_z=693(8)\,\mathrm{kHz}$. For this measurement, an interaction time of 100\,$\mu\mathrm{s}$ was used for the Raman transition as well as a Doppler detuning of 15\,MHz below resonance. The carrier (spin-flip) transition occurred at $\nu_0 = 138.911984$\,GHz.

Compared to~\cite{cornejoOpticalStimulatedRamanSideband2023}, where a lower axial frequency of 435\,kHz and a larger Doppler detuning of 20\,MHz below resonance had been chosen, here a similar axial temperature is reached after Doppler cooling, but a lower initial phonon number. This facilitates ground state cooling for two reasons: the number of potentially involved sideband transitions is reduced, as well as the number of required pulses. Working with this new parameter set only became possible after improvements related to laser beam position stability and ion loading protocols. Using these parameters, a mean phonon number of $\bar{n}_z \approx k_B T_z/h\nu_z = 94(9)$ in the axial mode can be determined after Doppler cooling for a single \BePlus ion.  

Once the blue sidebands frequencies are determined, resolved-sideband cooling can be performed. Each motion-reducing blue sideband pulse is followed by a repumper pulse of duration  20\,$\mu\mathrm{s}$ to re-set the spin to $\left|\uparrow\right>$. The sideband interaction strength depends on the phonon number $n$ of the axial motion~\cite{winelandExperimentalIssuesCoherent1998}; in particular, the sideband interaction strength of all sideband transitions vanishes close to certain phonon numbers~\cite{winelandExperimentalIssuesCoherent1998} across the range of relevant phonon numbers for our value of $\bar{n}_z$. It is therefore not sufficient to apply only 1st order sideband transitions for cooling because motional state population would get stuck around phonon numbers where the first order sideband interaction strength vanishes. 

We therefore start by applying alternating 6th and 5th order blue sideband pulses. The application of the sideband pulses in alternating order is favorable because the odd order sideband interaction strength tends to vanish for phonon numbers where the even order sideband interaction strength is significant and vice versa~\cite{cornejoOpticalStimulatedRamanSideband2023}. After 20 alternating sideband pulse pairs on the 6th and 5th order blue sidebands, we apply two more sequences of 20 alternating sideband pulse pairs on the 4th / 3rd  and 2nd / 1st order blue sidebands each. It is found experimentally that the most robust results were obtained by applying all sideband pulses for the same interaction time. 

\begin{figure}[t]
	\centering
	\includegraphics[width=1.0\columnwidth]{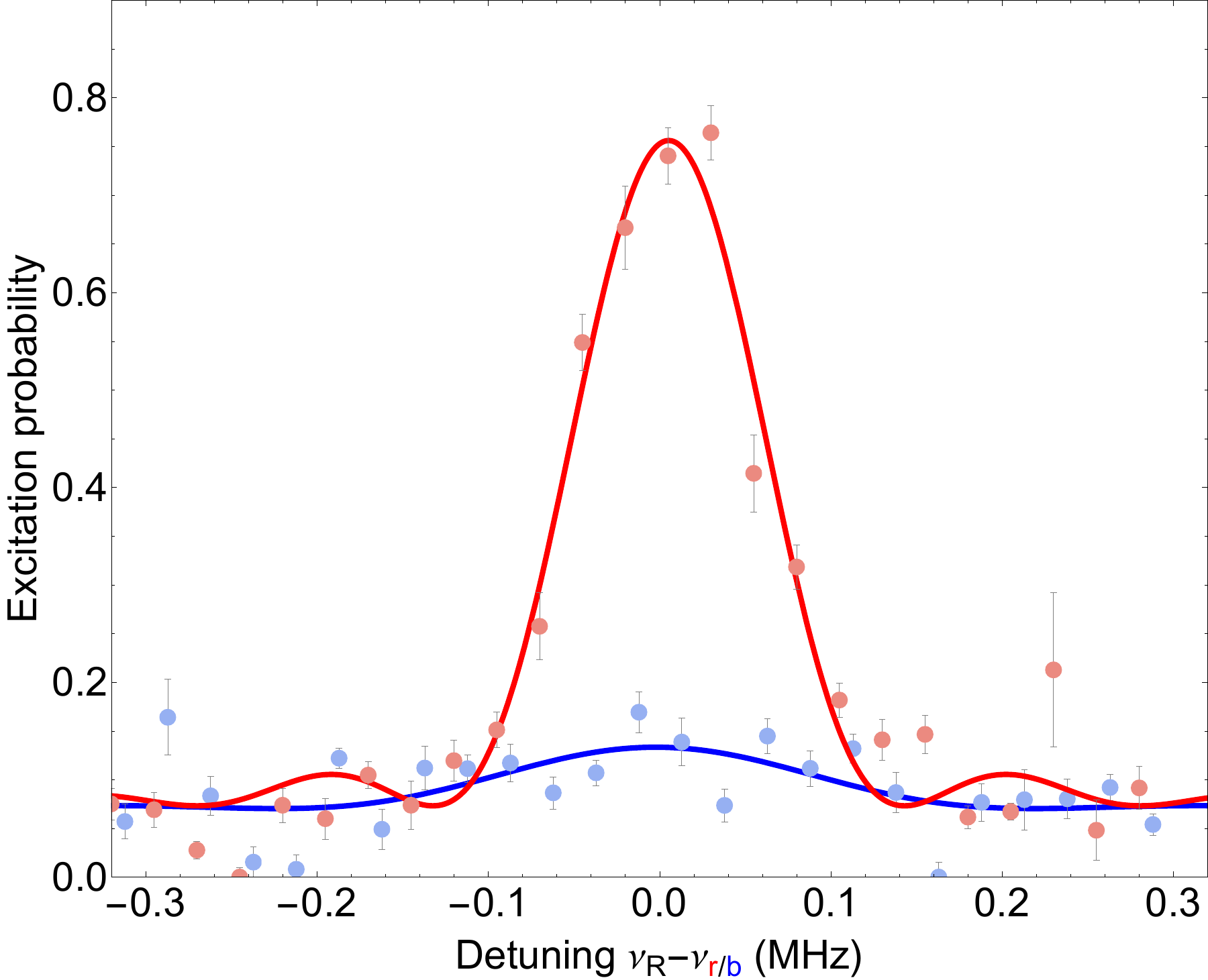}
	\caption{The excitation probability is shown as a function of the Raman laser beams' detuning. Data is depicted by blue and red dots for the first blue and red sidebands, respectively. Each data point and error bars are obtained from 100 measurements per data point. The blue and red lines are sinc-squared function fits to the data for the first blue and red sidebands. The center frequency of each transition is represented at zero detuning for clarity, where $\nu_{\mathrm{\textcolor{red}{r}/\textcolor{blue}{b}}}$ represents 138.911076~GHz and 138.912490~GHz for the first blue sideband and the first red sideband, respectively.}
	\label{fig:cooling}
\end{figure}  

The pulse sequence described above is able to cool a single \BePlus ion close to the motional ground state of the axial mode of motion. Figure~\ref{fig:cooling} shows the excitation probability of the first red and blue sideband ``analysis pulse'' after the cooling sequence. For the presented measurements, both beams are focused to a beam waist of around 150\,$\mu\mathrm{m}$ at the ion position and the laser power stabilized to 5\,mW and 1.7\,mW for Raman lasers 1 and 2, respectively. The difference in power is due to the required polarization of each Raman transition~\cite{mielke139GHzUV2021}. The sideband interaction time for each sideband cooling pulse was $10\,\mu\mathrm{s}$. The total of 120 sideband cooling pulses of $10\,\mu\mathrm{s}$ each and of 120 repump pulses of $20\,\mu\mathrm{s}$ each therefore took 3.6\,ms. The sideband interaction time for the red and blue sideband ``analysis pulse'' was $10\,\mu\mathrm{s}$. From the ratio of the maximum excitation probability of both sidebands (see Fig.~\ref{fig:cooling}), a mean phonon number in the axial mode of $\bar{n}_z = 0.10(4)$ is obtained. 

\begin{figure}[t]
	\centering
	\includegraphics[width=1.0\columnwidth]{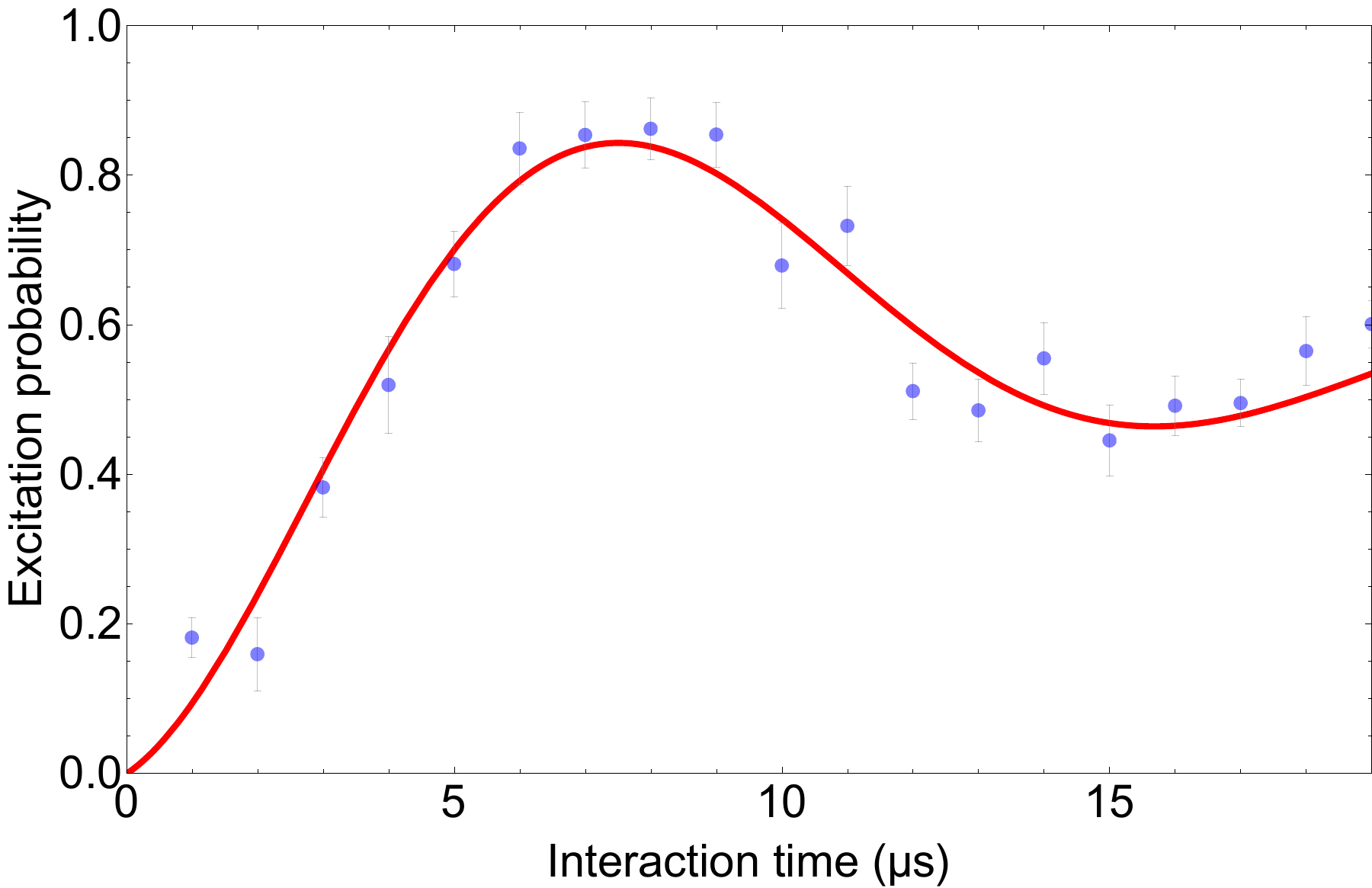}
	\caption{The excitation probability is shown as a function of the interaction time. Data are depicted by blue dots. Data and error bars are obtained from 100 measurements. The red line is a fit to the data using a sinusoidal exponential decay.}
	\label{fig:rabi}
\end{figure}

In order to test our ability to perform coherent state manipulation, Rabi oscillations on the first red sideband after Doppler and sideband cooling were analyzed. Figure~\ref{fig:rabi} shows the excitation probability for different interaction times of the red sideband ``analysis pulse'' on resonance with the first red sideband. A sinusoidal exponential decay is fitted to the data, yielding a frequency of 61(2)\,kHz and a decay time of 10(2)\,$\mu\mathrm{s}$. We assume that small variations in position of the Raman laser beams are the main causes of decoherence. This issue will be addressed by an active position stabilization system for both Raman laser beams.

\begin{figure}[b]
	\centering
	\includegraphics[width=1.0\columnwidth]{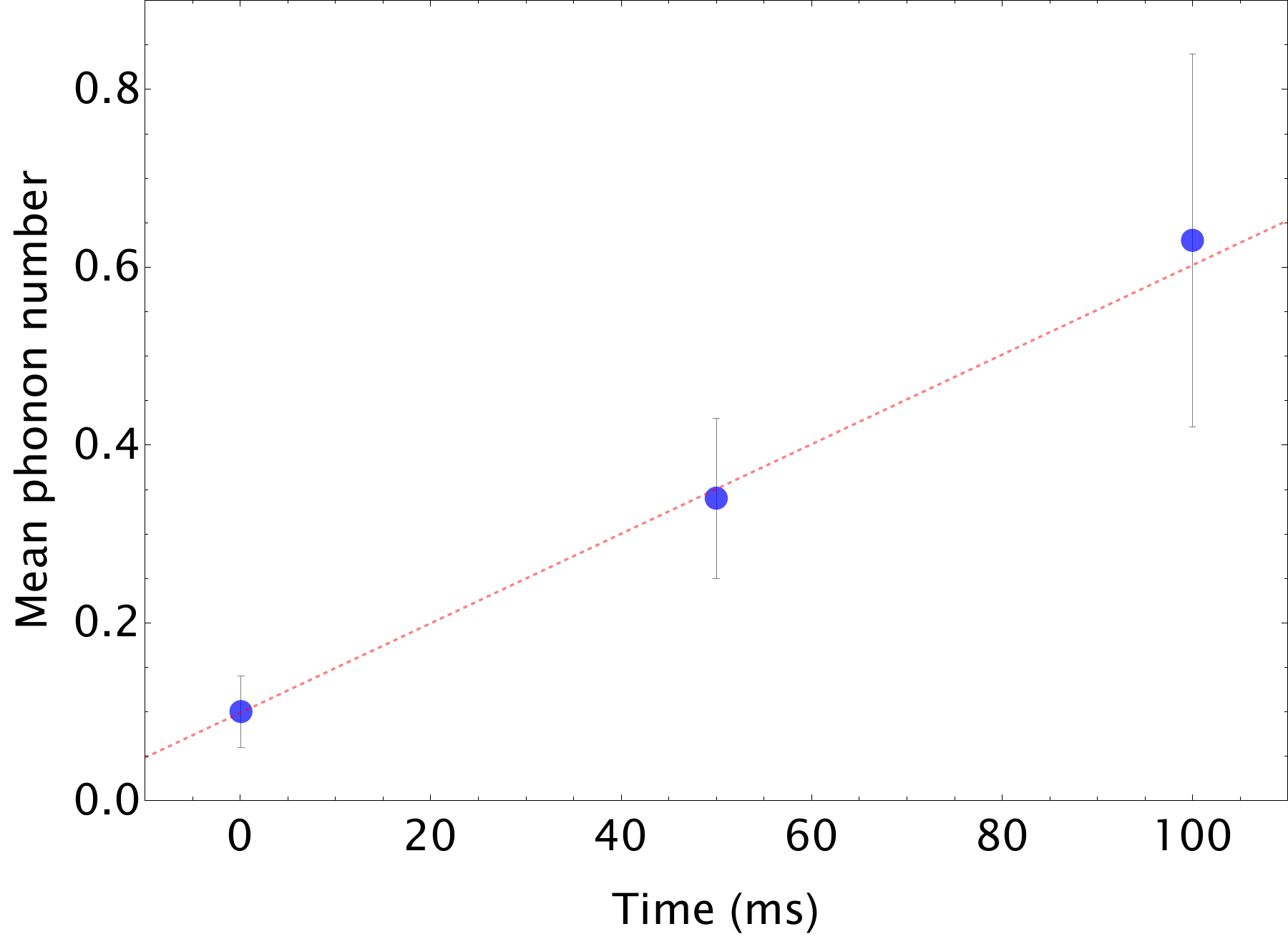}
	\caption{Heating rate measurement of a single laser-cooled \BePlus ion in our Penning trap system. The mean phonon number is measured after sideband cooling for a delay time of 10~$\mu\mathrm{s}$, 50~ms and 100~ms. Data are depicted by blue dots. The dashed red line is a linear fit to the data.}
	\label{fig:heating}
\end{figure}

One important aspect for the implementation of quantum logic spectroscopy in this system is to examine the rate at which the axial motion gains energy when otherwise left alone. We measure this key quantity by recording first blue and red sideband spectra following a variable delay time introduced after sideband cooling. Figure~\ref{fig:heating} shows the mean phonon number as a function of this delay time. From the linear fit to the data in Fig.~\ref{fig:heating}, a heating rate of $\dot{n}_z = (5.0 \pm 0.3)$~quanta/s is obtained. This corresponds to a noise spectral density scaled by the trap frequency of $\omega_{z}S_{E}(\omega_z) = 4m\hbar\omega_z^2\dot{n}_z/q^2$ = $(2.3 \pm 0.1)\times10^{-8}~\mathrm{V/m^2}$, where $\omega_z = 2\pi\nu_z$ and $\hbar$ is the reduced Planck constant. While still limited by technical noise, e.g. on the voltage sources for the electrodes, this measurement shows that the heating rate should already be low enough to consider the implementation of quantum logic spectroscopy, as all required steps for this are expected to happen on much shorter time scales. 

The results shown in this letter are an essential step towards the implementation of quantum logic spectroscopy~\cite{schmidtSpectroscopyUsingQuantum2005} in Penning traps~\cite{cornejoQuantumLogicInspired2021,nitzschkeElementaryLaserLessQuantum2020} for tests of CPT symmetry in the baryonic sector of the standard model. Because CPT and Lorentz symmetry are closely related, an earth-based experiment would be expected to measure sidereal variations of the observables\cite{bluhmCPTLorentzTests1998, kosteleckyDataTablesLorentz2011, dingLorentzCPTTests2019}. Such measurements would require sampling rates and accuracies that are difficult to imagine based on state-of-the-art resonator-based cooling techniques. The increased sampling rate and accuracy projected as a result of ground state cooling and quantum logic spectroscopy would enable such effects to be probed~\cite{cornejoQuantumLogicInspired2021}. More generally, the introduction of ground state cooling for precision measurements in Penning traps will enable such measurements to ultimately operate at the quantum limit. 

\begin{acknowledgments}
We acknowledge financial support from DFG through SFB/CRC  1227 ‘DQ-mat’, project B06 and through the cluster of excellence QuantumFrontiers, from the RIKEN Chief Scientist Program, RIKEN Pioneering Project Funding, the Max Planck-RIKEN-PTB Center for Time, Constants, and Fundamental Symmetries, and the European Research Council (ERC) under FP7 (grant Agreement No. 337154).
\end{acknowledgments}

\bibliography{qc}

\end{document}